\documentstyle[aps,pra,graphicx,float]{revtex}

\newcommand{\ket}[1]{\left|\mathrm{#1}\right\rangle}
\newcommand{\bra}[1]{\left\langle\mathrm{#1}\right|}
\begin{document}
\twocolumn[\hsize\textwidth\columnwidth\hsize\csname
@twocolumnfalse\endcsname

\title{Experimental Four-photon Entanglement and High-fidelity Teleportation}

\author{Jian-Wei Pan, Matthew Daniell,\protect\footnote{Present address: Department of Electrical and
        Computer Engineering, Boston University, 8 Saint Mary's Street,
        Boston, MA 02215} Sara Gasparoni, Gregor Weihs, and Anton Zeilinger}
\address{Institut f\"ur Experimentalphysik, Universit\"at Wien, Boltzmanngasse 5, 1090 Wien, Austria}
\date{\today}

\maketitle

\begin{abstract}
We experimentally demonstrate observation of highly pure
four-photon GHZ entanglement produced by parametric
down-conversion and a projective measurement. At the same time
this also demonstrates teleportation of entanglement with very
high purity. Not only does the achieved high visibility enable
various novel tests of quantum nonlocality, it also opens the
possibility to experimentally investigate various quantum
computation and communication schemes with linear optics. Our
technique can in principle be used to produce entanglement of
arbitrarily high order or, equivalently, teleportation and
entanglement swapping over multiple stages.
\end{abstract}


\vskip1pc]

\narrowtext

Entanglement is not only the essence of quantum mechanics as
suggested by Erwin Schr{\"o}dinger \cite{schrodinger35a}, but is
also at the basis of nearly all quantum information protocols such
as quantum cryptography, quantum teleportation and quantum
computation\cite{Bouwmeester00a}. While entanglement of two qubits
is routine in the laboratory, entanglement of three photons
\cite{Greenberger89a} with high quality has only recently been
experimentally realized \cite{Bouwmeester98a} and used to
experimentally demonstrate the extreme contradiction between local
realism and quantum mechanics \cite{Pan00a} in so-called GHZ
states. In a parallel development entanglement of the quantum
states of three atoms \cite{Rauschenbeutel00a} or four qubits in
ions \cite{Sackett00a} has been demonstrated, yet in all these
cases the quality of the entangled states still needs to be
significantly improved in order to be useful for tests of quantum
mechanics or in quantum information schemes.

A similar situation is found in the recent teleportation
experiments \cite{Bouwmeester97a,Boschi98a,Pan98a,Furusawa98a}. To
verify the nonlocal character of teleportation, two conditions
must be satisfied in any experiment. On the one hand, one has to
demonstrate that a genuinely unknown state (in the optimal case, a
qubit which itself is still entangled to another one) is
teleported \cite{Bennett93a}, on the other hand a high
experimental visibility is necessary in order to exclude local
hidden variable models
(LHV)\cite{Zukowski00a,Grangier00a,Clifton01a,Barrett01a}. The
so-called entanglement swapping experiment \cite{Pan98a} is the
only one to date that demonstrates the teleportation of a
genuinely unknown state. However, since its observed visibility
was lower than $71\%$, one could in principle still doubt the
nonlocal feature of teleportation\cite{Zukowski00a}.

In this letter we report on an experiment that not only
demonstrates the observation of four-photon entanglement but also
shows high-fidelity entanglement swapping, thus proving the
nonlocal character of quantum teleportation. Both features are not
only important for performing novel fundamental experiments to
test quantum mechanics or to demonstrate its counter-intuitive
features, but also to expand our toolbox for quantum computation
and quantum communication.

\begin{figure}[ht]
\begin{center}
\includegraphics[width=0.5\columnwidth]{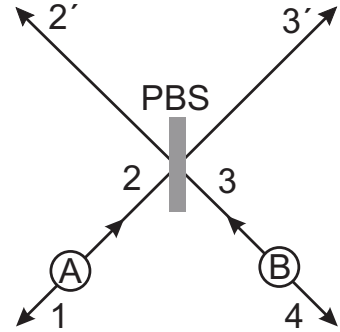}
\end{center}
\caption{Principle for observing four-photon GHZ correlations.
Sources A and B each deliver one entangled particle pair. A
polarizing beam-splitter (PBS) combines modes 2 and 3. The two
photons detected one each in its output port are either both H
(horizontally) or both V (vertically) polarized projecting the
complete four-photon state into a GHZ state.} \label{schematic}
\end{figure}

Our technique of observing four-photon GHZ entanglement uses two
independently created photon pairs (Fig.~\ref{schematic}). Suppose
that the two pairs are in the state
\begin{eqnarray}
    \label{s1}
    \left|\Psi^i\right\rangle_{1234}& = & \frac{1}{\sqrt{2}}
    \left( \left| \mbox{H} \right\rangle_{1} \left|\mbox{V}
    \right\rangle_{2} - \left| \mbox{V} \right\rangle_{1} \left|
    \mbox{H} \right\rangle_{2} \right) \otimes \nonumber \\
    & &  \frac{1}{\sqrt{2}}
    \left( \left| \mbox{H} \right\rangle_{3} \left|\mbox{V}
    \right\rangle_{4} - \left| \mbox{V} \right\rangle_{3} \left|
    \mbox{H} \right\rangle_{4} \right),
\end{eqnarray}
which is a tensor product of two polarization entangled photon
pairs. Here $\left| \mbox{H} \right\rangle$  ($\left|
\mbox{V}\right\rangle$) indicates the state of a horizontally
(vertically) polarized photon.

One photon out of each pair is directed to the two inputs of a
polarizing beam-splitter (PBS). Since the PBS transmits horizontal
and reflects vertical polarization, coincidence detection between
the two PBS outputs implies that either both photons 2 and 3 are
horizontally polarized or both vertically polarized, and thus
projects the state (1) onto a two-dimensional subspace spanned by
$\ket{V}_1\ket{H}_2\ket{H}_3\ket{V}_4$ and $
\ket{H}_1\ket{V}_2\ket{V}_3\ket{H}_4$.

After the PBS, the state corresponding to a four-fold coincidence
is
\begin{equation}
\label{s2} \left|\Psi^f\right\rangle_{12^{\prime}3^{\prime}4} =
\frac{1}{\sqrt{2}} \left( \left| \mbox{H} \right\rangle_{1}
\left|\mbox{V} \right\rangle_{2^{\prime}}\left| \mbox{V}
\right\rangle_{3^{\prime}} \left| \mbox{H} \right\rangle_{4} +
\left| \mbox{V} \right\rangle_{1} \left|\mbox{H}
\right\rangle_{2^{\prime}}\left| \mbox{H}
\right\rangle_{3^{\prime}} \left| \mbox{V} \right\rangle_{4}
\right).
\end{equation}
This is a GHZ-state of four particles, which can exhibit nonlocal
behavior according to the GHZ theorem.

The scheme described above has several notable features. First, it
yields a four-fold coincidence with a success probability of
$50\%$, which is much (four times) more efficient than the one
reported for observation of three-photon entanglement
\cite{Bouwmeester98a}. Second, the scheme does not only yield
four-particle entanglement but --- assuming perfect pair sources
and detectors --- could also produce freely propagating
three-particle entangled states of modes 1, $3^{\prime}$, and 4,
if one puts a $45^\circ$ polarizer into output $2^{\prime}$.
Detecting one photon in one of the outputs of this polarizer makes
sure that there will be exactly one photon in each of the outputs
1, $3^{\prime}$, and 4. Finally, by this technique one can also
implement teleportation of entanglement, and hence a realization
of entangled pair production with event-ready
detectors\cite{Zukowski93a}. To do this two $45^\circ$ polarizers
are inserted into outputs $2^{\prime}$ and $3^{\prime}$.
Conditioned on a coincidence detection of one photon in each of
these outputs, we obtain an entangled pair in outputs 1 and 4 (for
more details, see our further discussion below). Note that we can
notify the observers at 1 and 4 before their measurements.

\begin{figure}[ht]
\begin{center}
\includegraphics[width=0.70\columnwidth]{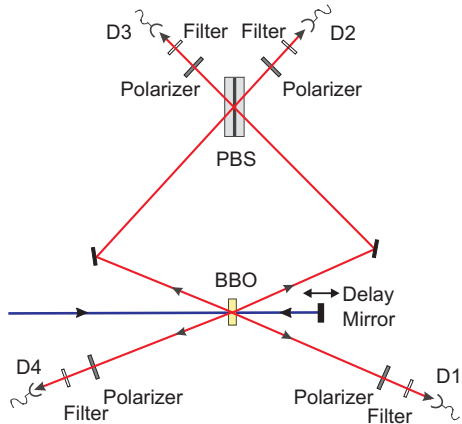}
\end{center}
\caption{Schematic of the experimental setup for the measurement
of four-photon GHZ correlations. A pulse of UV-light passes a BBO
crystal twice to produce two entangled photon pairs. Coincidences
between all four detectors 1-4 exhibit GHZ entanglement.}
\label{setup}
\end{figure}

Obviously, an optimal realization of the above scheme would
require perfect photon pair sources and ultimately perfect
single-photon detectors. However, it is important to note that the
absence of perfect sources and detectors does not prevent us from
performing an experimental demonstration because, on the one hand,
any practical application of our scheme would always need a final
verification step by detecting a four-fold coincidence. On the
other hand, any method to ensure sources A and B each emit one
only one entangled pair is in essence equivalent to a four-fold
coincidence detection. In the following we are going to describe
our experimental verification of four-photon GHZ correlations.

In our experiment (see Fig.~\ref{setup}) we create
polarization-entangled photon pairs by spontaneous parametric
down-conversion from an ultraviolet femtosecond pulsed laser
($\sim 200$fs, $\lambda \simeq 394.5$nm) in a
$\beta-\mathrm{BaB_3O_6}$(BBO) crystal
\cite{Bouwmeester97a,Kwiat95b}. The laser passes the crystal a
second time having been reflected off a translatable mirror. In
the reverse pass another conversion process may happen producing
an second entangled pair. One particle of each pair is steered to
a polarizing beam splitter where the path lengths of each particle
have been adjusted such that they arrive simultaneously. On the
polarizing beam-splitter a horizontally polarized photon will
always be transmitted whereas a vertically polarized one will
always be reflected both with less than $10^{-3}$ error rate. The
two outputs of the polarizing beam-splitter are spectrally
filtered (3.5~nm bandwidth) and monitored by fiber-coupled single
photon counters (D2 and D3). The filtering process stretches the
coherent time to about 550fs, substantially larger than the pump
pulse duration\cite{Zukowski95a}. This effectively erases any
possibility to distinguish the two photons according to their
arrival time and therefore leads to interference.

The remaining two photons -- one from each pair -- pass identical
filters in front of detectors D1 and D4 and are detected directly
afterwards. In front of each of the four detectors we may insert a
polarizer to assess the correlations with respect to various
combinations of polarizer orientations. A correlation circuit
extracts only those events where all four detectors registered a
photon within a small time window of a few ns. This is necessary
in order to exclude cases in which only one pair is created or two
pairs in one pass of the pump pulse and none in the other.

\begin{figure}[ht]
\begin{center}
\includegraphics[width=1.00\columnwidth]{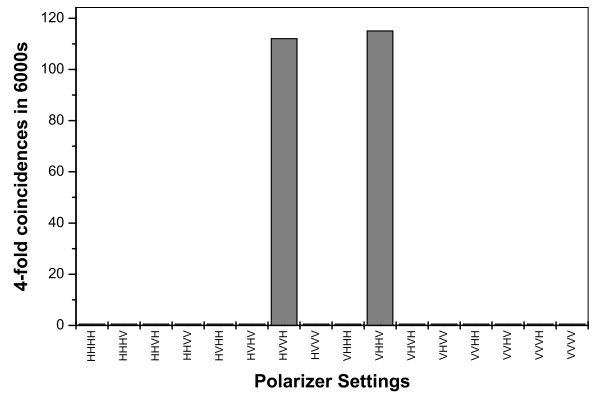}
\end{center}
\caption{Experimental data for horizontal and vertical polarizer
settings. Only the two desired terms are present; all other terms
which are not part of the state $\ket{\Psi^f}$
(Eq.~(\protect\ref{s2})) are so strongly suppressed that they can
hardly be discerned in the graph. The number of four-fold
coincidences for any of the non-desired terms is 0.5 in 6000s on
the average.} \label{datahv}
\end{figure}

To experimentally demonstrate that the state $\ket{\Psi^f}$ of
Eq.~(\ref{s2}) has been obtained, we first verified that under the
condition of having a four-fold coincidence only the HVVH and VHHV
components can be observed, but no others. This was done by
comparing the count rates of all sixteen possible polarization
combinations, HHHH, ... VVVV. The measurement results in the H/V
basis (Fig.~\ref{datahv}) show that the signal to noise ratio
defined as the ratio of any of the desired four-fold events (HVVH and
VHHV) to any of the 14 other non-desired ones is about 200:1.

Showing the existence of HVVH and VHHV terms alone is just a
necessary but not a sufficient experimental criterion for the
verification of the state $\ket{\Psi^f}$, since the above
observation is in principle both compliant with $\ket{\Psi^f}$
and with a statistical mixture of HVVH and VHHV. Thus, as a
further test we have to demonstrate that the two terms HVVH and
VHHV are indeed in a coherent superposition.

\begin{figure}[ht]
\begin{center}
\includegraphics[width=1.00\columnwidth]{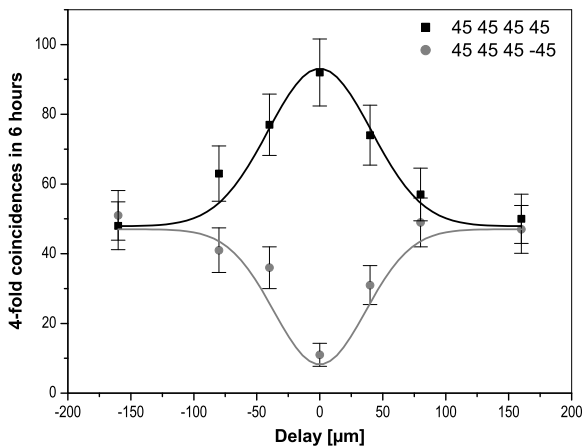}
\end{center}
\caption{Experimental data for $45^\circ$ polarizer settings. The
difference between the fourfold coincidence count rates for
($+45^\circ$/$+45^\circ$/$+45^\circ$/$+45^\circ$) and
($+45^\circ$/$+45^\circ$/$+45^\circ$/$-45^\circ$) shows that the
amplitudes depicted in Fig.~\protect\ref{datahv} are in a coherent
superposition. Maximum interference occurs at zero delay between
the photons 2 and 3 arriving at the polarizing beam-splitter. The
gaussian curves that roughly connect the data points are only
shown to guide the eye. Visibilities and errors are calculated
only from the raw data.} \label{data45}
\end{figure}

This was done by further performing a  polarization measurement in
the $45^{\circ}$ basis, where
$\ket{+45^{\circ}}=1/\sqrt{2}(\ket{H}+\ket{V})$ and
$\ket{-45^{\circ}}=1/\sqrt{2}(\ket{H}-\ket{V})$ are two
corresponding eigenstates. Transforming $\ket{\Psi^f}$ to the
$\ket{+45^{\circ}},\ket{-45^{\circ}}$ linear polarization basis
yields an expression containing eight (out of 16 possible) terms,
each with an even number of $\ket{+45^{\circ}}$ components.
Combinations with odd numbers of $\ket{+45^{\circ}}$ components do
not occur. As a test for coherence we can now check the presence
or absence of various components. In Fig.~\ref{data45} we compare
the ($+45^\circ$/$+45^\circ$/$+45^\circ$/$+45^\circ$) and
($+45^\circ$/$+45^\circ$/$+45^\circ$/$-45^\circ$) count rates as a
function of the pump delay mirror position. At zero delay --
photons 2 and 3 arrive at the PBS simultaneously -- the latter
component is suppressed with a visibility of $0.79 \pm 0.06$. As
explained in Ref.~\cite{Zukowski95a}, many efforts have been made
by us to obtain this high visibility reliably. In the experiment
we observed that the most important ingredients for a high
interference contrast were a high single pair entanglement
quality, the use of narrow bandwidth filters, and the high quality
of the polarizing beam-splitter.

These measurements clearly show that we obtained four-particle GHZ
correlations. The quality of the correlations can be judged by
the density matrix of the state
\begin{equation}
 \rho=0.89 \ket{\Psi^f}\bra{\Psi^f}_{12^{\prime}3^{\prime}4} +
      0.11 \ket{\Phi}\bra{\Phi}_{12^{\prime}3^{\prime}4}
\end{equation}
where $\ket{\Phi}=1/\sqrt{2}(\ket{HVVH}-\ket{VHHV})$. This density
matrix describes our data under the experimentally well justified
assumption that only phase errors in the H/V basis are present,
which appear as bit-flip errors in the $45^\circ$ basis (see
Fig.~\ref{data45}).

To show our experiment is also a realization of entanglement
swapping, let us rewrite the state of Eq.~(1) in the following way
\begin{eqnarray}
    \label{s4}
    \left|\Psi^i\right\rangle_{1234}& = & \frac{1}{2}
    \left(\left| \psi^{+} \right\rangle_{14}
    \left| \psi^{+} \right\rangle_{23} -
    \left| \psi^{-} \right\rangle_{14}
    \left| \psi^{-} \right\rangle_{23}
    \right. \nonumber \\
    & & \left. - \left| \phi^{+} \right\rangle_{14}
    \left| \phi^{+} \right\rangle_{23} +
    \left| \phi^{-} \right\rangle_{14}
    \left| \phi^{-} \right\rangle_{23}
    \right),
\end{eqnarray}
where
\begin{equation}
\label{s5}
\begin{array}{l}
\left| \psi^{\pm} \right\rangle = \frac{1}{\sqrt{2}}(\ket{H}\ket{V}
\pm \ket{V}\ket{H}),\\ \left| \phi^{\pm} \right\rangle =
\frac{1}{\sqrt{2}}(\ket{H}\ket{H} \pm \ket{V}\ket{V})
\end{array}
\end{equation}
are the four orthogonal Bell states.

Suppose that we now perform a joint Bell-state measurement on
photons 2 and 3, i.e. project photons 2 and 3 onto one of the four
Bell states. Eq.~(4) implies that this measurement also
correspondingly projects photons 1 and 4 onto the same Bell state.
After projection of photons 2 and 3, in all four cases photons 1
and 4 emerge entangled although they never interacted with one
another in the past. This is the so-called entanglement
swapping\cite{Zukowski93a}, which can also be seen as
teleportation of either of the state of photon 2 over to photon 4
or the state of photon 3 over to photon 1\cite{Bennett93a}.
Apparently, in order to experimentally show the working principle
of entanglement swapping it is sufficient to identify only one of
the four Bell states\cite{Bouwmeester97a,Bouwmeester00b}.

In the experiment, we chose to analyze the projection onto $\left|
\phi^+ \right\rangle_{23}$. This projection is accomplished by
performing a polarization decomposition in the $45^{\circ}$ basis
in outputs $2^{\prime}$ and $3^{\prime}$ and a subsequent
coincidence detection\cite{Pan98b}. More explicitly, detecting
$+45^\circ$/$+45^\circ$ or $-45^\circ$/$-45^\circ$ coincidences
between the outputs $2^{\prime}$ and $3^{\prime}$ acts as a
projection onto $\left| \phi^+ \right\rangle_{23}$, and thus
leaves photons 1 and 4 in the identical state $\left| \phi^+
\right\rangle_{14}$. This behavior is verified by the data shown
in Figs.~\ref{datahv},\ref{data45}. Fig.~\ref{datahv} proves that
only HH and VV terms are present in the state of particles 1 and 4
conditioned on a fourfold coincidence. Fig.~\ref{data45} in turn
can be viewed as the interference pattern showing the correlation
in the conjugate basis. Specifically, the data of
Fig.~\ref{data45} indicate that the state of, say, photon 2 was
teleported to photon 4 with a fidelity of 0.89. This clearly
outperforms our earlier work \cite{Pan98a} in this field, and for
the first time fully demonstrates the nonlocal feature of quantum
teleportation\cite{Zukowski00a}.

An experimental realization of four-particle GHZ entanglement and
high-fidelity teleportation has rather profound implications.
First, going to higher entangled systems the contradiction with
local realism becomes ever stronger, because both the necessary
visibility and the required number of statistical tests to reject
the LHV models at a certain confidence level decrease with the
number of particles that are entangled\cite{Zukowski97a,Peres00a}.
Second, based on the observed visibility of $0.79 \pm 0.6$ one
could violate -- with an appropriate set of polarization
correlation measurements -- Bell's inequality for photons 1 and 4,
even though these two photons never interacted directly. As noted
by Aspect, "This would certainly help us to further understand
nonlocality" \cite{Collins98a}. In our experiment however, due to
the low count rates and some instability in the pump laser it was
not yet possible to carry out all the measurements needed. Note
that, with the present experimental performance a continuous
measurement of more than six months would be necessary to collect
statistically sufficient data.

Besides its significance in tests of quantum mechanics versus
local realism, the methods developed in the experiment also have
many useful applications in the field of quantum information. It
was noticed very recently that while our setup directly provides a
simple way to perform entanglement concentration
\cite{Yamamoto01a,Zhao01a}, a slight modification of the setup
also provides a novel way to perform entanglement purification for
general mixed entangled states\cite{Pan01a}. Furthermore,
following the recent proposal by Knill et al.\cite{Knill01a}, our
four-photon experiment also opens the possibility to
experimentally investigate the basic elements of quantum
computation with linear optics.

In summary, we have demonstrated a method of creating higher order
entangled states which can in principle be extended to any desired
number of particles, provided one has efficient pair sources.
Given that, more photon pair sources could be combined with
polarizing beam-splitters to yield entangled states of arbitrary
numbers of particles. Latest developments in photon pair sources
suggest that it should be possible in the near future to have
sources with many orders of magnitude higher emission rates
\cite{Sanaka00a,Tanzilli01a}. With these entanglement sources one
would be able to implement some quantum computation algorithms
using only entanglement and linear optics \cite{Knill01a}. Also,
more elaborate entanglement purification protocols and
high-fidelity teleportation over multiple stages as required for
the construction of quantum repeaters\cite{Briegel98a} become
possible.

We acknowledge the financial support of the Austrian Science Fund
FWF, project no. F1506 and the European Commission within the
IST-FET project ``QuComm'' and TMR network ``The physics of
quantum information.''

\end{document}